\documentclass[10pt]{iopart}
\usepackage{iopams}
\usepackage{graphicx}% Include figure files
\usepackage{bm}% bold math
\usepackage{iopams,setstack}
\usepackage{graphicx}% Include figure files

\newcommand{\beq}{\begin{equation}}
\newcommand{\eeq}{\end{equation}}
\newcommand{\beqa}{\begin{eqnarray}}
\newcommand{\eeqa}{\end{eqnarray}}
\def\nn{\nonumber\\}
\def\eq#1{(\ref{#1})}

\def\cd#1{\ensuremath{\nabla_{#1}}}          %covariant derivative
\def\pd#1{\ensuremath{\partial_{#1}}}        %partial derivative
\def\st{spacetime}

\def\mch{\scriptscriptstyle}

\def\text#1{{\rm #1}}

\def\lab#1{\label{#1}}
\def\hz{\hat{z}}

\begin{document}

\title{{Static plane symmetric relativistic fluids and empty repelling singular boundaries }}

\author{Ricardo\ E.  Gamboa Sarav\'{\i}\dag\  \ddag
}

\address{\dag\ Departamento de F\'{\i}sica, Facultad de Ciencias
Exactas, Universidad Nacional de La Plata, Argentina. }
\address{\ddag\ IFLP, CONICET,
Argentina.  }

\ead{quique@fisica.unlp.edu.ar}

\date{\today}

%%%%%%%%%%%%%%%%%%%%%%%%%%%%%%%%%%%%%%%%%%%%%%%%%%%%%%%%%%%%%%%%%%%%%%%
\begin{abstract}
We present a  detailed analysis of the general exact solution of Einstein's equation
corresponding to a static and plane symmetric  distribution of
 matter with density proportional to pressure. We study the
geodesics in it and we show that this simple spacetime exhibits
very curious  properties. In particular, it has a free of matter repelling singular boundary
and all geodesics bounce off it.
\end{abstract}

\submitto{\CQG} 

%%%%%%%%%%%%%%%%%%%%%%%%%%%%%%%%%%%%%%%%%%%%%%%%%%%%%%%%%%%%%%%%%%%%%%%%%%%%%%%%%%%%%%%%%%%%

\section{Introduction}

Because of  the complexity of Einstein's field equations, one
cannot find exact solutions except in spaces of rather high
symmetry---very often with no direct physical application.
Nevertheless, exact solutions can give an idea of the qualitative
features that could arise in General Relativity  and so, of possible
properties of realistic solutions of the field equations.

In this paper we want to illustrate some curious features of gravitation by means of a simple  solution:  the gravitational field of a static plane symmetric relativistic perfect fluid.

We have recently  pointed out that solutions of  Einstein's equation
presenting an empty (free of matter) repelling boundary where spacetime curvature diverges
occur \cite{gs}, and we called this kind of singularities {\em
white walls.} These  singularities are not the sources of the fields but they arise owing to  the attraction of distant matter.

The solution we described in \cite{gs} is the
gravitational field  of a static homogenous distribution of matter with plane symmetry lying below $z=0$. Because of the symmetry required,  the  exterior   gravitational field turns out to be Taub's plane vacuum solution \cite{taub}.

 Although the properties of Taub's plane solution have been known for
several years and perhaps due to the belief that  singularities are always the  sources of vacuum solutions, this solution has been usually associated to matter with negative mass
(see \cite{bonnor,bed} and references therein).
For example, the authors of Ref. \cite{bed} suggest  that it must be interpreted as
representing the exterior gravitational field  of a planar shell with {\em infinite
negative} mass density  at the singularity, whereas we argue  that Taub's plane solution is also the external gravitational field of ordinary (i.e., with nonnegative density and pressure) matter sitting at negative values of $z$ and that the singularity that arises high above is due to the attraction of the  distant matter \cite{gs}.

Thus, it would be worth  finding an exact
solution of Einstein's equation corresponding to a  distribution of ordinary matter presenting  a singularity of this kind.

The aim of this paper is to show that the solution found by Collins \cite{collins}  for a static and plane symmetric  relativistic perfect
fluid obeying an equation of state such that $\rho$ and $p$ are proportional to each other (see also \cite{hojman} and  for the case $\rho =p$ \cite{romualdo}) exhibits such a property.

 In Sec. II we present a simple and direct  derivation  of Collins's solution, as well as a detailed analysis of it.   In Sec. III we study the geodesics in it.

Throughout this paper, we adopt the convention in which the \st\
metric has signature $(-\ +\ +\ +)$, the  system of units in which
the speed of light $c=1$ and Newton's gravitational
constant $ G=1$.

\section{The Collins solution}

We want to find  the solution  of Einstein's
equation corresponding to  a static and plane symmetric
distribution of matter with plane symmetry. That is, it must be
invariant under translations in the plane and under rotations
around its normal. The matter we will consider is  a perfect fluid
satisfying the equation of state \footnote{Collins in Ref. \cite{collins}  uses $p=(\gamma-1)\rho$, so $\eta=(\gamma-1)^{-1}$.}\beq\lab{ee} \rho= \eta\,
p\,,\eeq where $\eta$ is an arbitrary constant---for ordinary
matter $0<p<\rho$ and so $\eta>1$. The stress-energy tensor is
\beq T_{ab}= (\rho+p)\,u_au_b+p\, g_{ab}\, ,\eeq where $u^a$ is
the velocity of fluid elements.

 Due to the plane symmetry and staticity, following \cite{taub} we can find coordinates $(t, x, y, z)$ such that \beq
\lab {met} ds^2= - e^{2 U(z)}\left(dt^2-dz^2\right)+
e^{2V(z)}\left(dx^2+dy^2
\right)\,.\eeq%
That is, it is the more general metric admitting the Killing vectors $\pd x$, $\pd y$, $x\pd y-y\pd x$ and $\pd t$.

The non-identically vanishing components of the Einstein tensor are
\beqa \lab {gtt} G_{tt}=2\ \pd z U\, \pd zV- 3\ (\pd z V)^2 - 2\,
\pd z \pd z V, \\
\lab {gii} G_{xx}=G_{yy}= e^{-2U+2V} \left((\pd zV)^2 + \pd z \pd
z V +\pd z \pd z U \right)\,,
\\ \lab {gzz} G_{zz}= \pd z  V\left(2\ \pd z U+ \pd z V\ \right) .
 \eeqa%
On the other hand, since the fluid must be static,
$u^a=(e^{-U},0,0,0)$,  so \beq T_{ab}= \text{diag}\left(\rho\,
e^{2U},p\, e^{2V},p\, e^{2V},p\, e^{2U}\right)\,, \eeq where
$\rho$ and $p$ can depend only on the z-coordinante.
 Thus, Einstein's equations, i.e., $G_{ab}=8 \pi T_{ab}$, are
\beqa \lab {gtt1} 2\ \pd z U\, \pd zV- 3\ (\pd z V)^2 - 2\, \pd z
\pd z V= \tilde{\rho}\,e^{2U}, \\%
\lab {gii1} (\pd zV)^2 + \pd z \pd z V +\pd z \pd z U =
\tilde{p}\,e^{2U} ,\\
\lab {gzz1}   \pd z  V\left(\ 2\ \pd z U+ \pd z V\ \right)=
\tilde{p}\,e^{2U}\,,
 \eeqa%
where $\tilde{\rho}= 8 \pi \rho$ and $\tilde{p}= 8 \pi p$.

On the other hand, $\cd a T^{ab}=0$ yields
\beq \lab{ppr} \pd z \tilde{p} = -(\tilde{p}+\tilde{\rho})\,\pd z
U.\eeq Of course, due to Bianchi's identities, equations
(\ref{gtt1}), (\ref{gii1}), (\ref{gzz1}) and (\ref{ppr}) are not
independent, so we will use  only \eq{gtt1}, \eq{gzz1}, and
\eq{ppr}.

By using the equation of state \beq \lab{eqs}\rho(z)= \eta\ p(z)\, ,
\eeq we can write \eq {ppr} as
\beq \pd z\left(\ln \tilde{p} +( \eta+1) U\right)=0 \eeq%
and so
 \beq\lab{presion}\tilde{p} = C\,
e^{-(\eta+1)U},  \eeq where $C$ is an arbitrary constant. Thus,
\eq{gtt1} and \eq{gzz1} become
\beqa \lab {gtt2} 2\ \pd z U\, \pd zV- 3\ (\pd z V)^2 - 2\, \pd z
\pd z V= \eta\, C \,e^{(1-\eta)U}, \\ \lab {gzz2}  \pd z  V\left(\ 2\ \pd z U+ \pd z V\ \right)=  C \,e^{(1-\eta)U}. \eeqa%
The change of variables \beq \lab{xi} \xi = \int_{z_0}^{z}
e^{(1-\eta)U(u)/2}du, \eeq brings \eq{gtt2} and \eq{gzz2} to
\beqa \lab {gtt3} (1+\eta)\ \pd \xi U\, \pd \xi V- 3\ (\pd \xi
V)^2 - 2\, \pd \xi \pd \xi V= \eta\, C , \\
\lab {gzz3}  2\ \pd \xi U\, \pd \xi  V + (\pd \xi V)^2 = C.  \eeqa%
Hence
\beq \lab {gtt4} \pd \xi \pd \xi V + \frac{(\eta+7)}{4}\ (\pd \xi V)^2 +\frac{(\eta-1)}{4}\ C =0. \eeq%

By writing \beq \lab{W}W(\xi)= e^{\frac{\eta+7}{4}V(\xi)},\eeq we
can write Eq. \eq{gtt4} as
\beq\lab {gtt5} \pd \xi \pd \xi W= -
\frac{C(\eta+7)(\eta-1)}{16}\,W\,.\eeq
Setting ${\alpha}=C (\eta+7)(\eta-1)/16$, we can write the general
solution of \eq{gtt5} as%
 \beqa \lab{W1} W(\xi)= \cases{C_1\, \frac{\sin\sqrt{\alpha}(\xi+C_2)}{\sqrt{\alpha}}&{if}\,\,\,\ $\alpha>0$\\
 C_1\, \xi+C_1\,C_2&{if}\,\,\,\ $\alpha=0$\\C_1\, \frac{\sinh\sqrt{-\alpha}(\xi+C_2)}{\sqrt{-\alpha}}&{if}\,\,\,\ $\alpha<0$\ ,}\eeqa %
where $C_1$ and $C_2$ are arbitrary constants. Notice that the
three cases in \eq{W1} are covered (as a limit when $\alpha=0$)
by the first one. Therefore, taking into account \eq{W}, we can
write
for arbitrary $\alpha$%
 \beq V(\xi)=\ln\left(
\frac{C_1\,\sin\sqrt{\alpha}(\xi+C_2)}{\sqrt{\alpha}}\right)^{\frac{4}{\eta+7}}.\eeq
Now, from \eq{gzz3} we can write $\pd \xi U$ in terms of $V(\xi)$
as
\beq\lab{gzz4}   \pd \xi U\,   = \frac{ C }{2\,\pd \xi  V}- \frac{\pd \xi V}{2}.\eeq%
By integrating it, we find  the general solution for $U(\xi)$%
\beq \fl U(\xi)=
\,\ln\left(\left(\frac{C_3\,\sin\sqrt{\alpha}(\xi+C_2)}{\sqrt{\alpha}}\right)^{-\frac{2}{\eta+7}}{\left(\cos\sqrt{\alpha}(\xi+C_2)\right)}^{-\frac{2}{\eta-1}}\right),\eeq
where $C_3$ is a new arbitrary constant.%

% \beq e^{U(\xi)}=
%\left(\frac{C_3\,\sin\sqrt{\alpha}(\xi+C_2)}{\sqrt{\alpha}}\right)^{-\frac{2}{\eta+7}}\left({\cos\sqrt{\alpha}(\xi+C_2)}\right)^{-\frac{2}{\eta-1}}\eeq

If we now make the transformation: $(t,x,y,z)\rightarrow(t,x,y,\xi)$, the line element \eq{met} becomes%
\beqa \lab{met1}%\fl
ds^2=  -  \left(
 \frac{C_3\, \sin\sqrt{\alpha} (\xi+C_2)}{\sqrt{\alpha}}\right)^{-\frac{4}{\eta +7}}\,\Bigl(\cos\sqrt{\alpha}
(\xi+C_2)\Bigr)^{-\frac{4}{\eta-1}}\,
 dt^2 \nn + \left(\frac{C_1\,\sin\sqrt{\alpha} (\xi+C_2)}{\sqrt{\alpha}}\right)^{\frac{8}{\eta+7}}\left(dx^2+dy^2\right) \nn+\,\left(
 \frac{C_3\,\sin\sqrt{\alpha} (\xi+C_2)}{\sqrt{\alpha}}\right)^{-2\frac{\eta+1}{\eta +7}}\,\Bigl(\cos\sqrt{\alpha}
(\xi+C_2)\Bigr)^{-2\frac{\eta+1}{\eta-1}}\, d\xi^2,
 \eeqa%
  since from \eq{xi} we see that
\beq \lab{gxixi}g_{\xi\xi}=\left(\pd \xi z\right)^2 g_{zz}= e^{(\eta-1)U} e^{2U}=e^{(\eta+1)U}. \eeq %

%where the range of $\xi$ is $0\leq\xi\ \leq \pi/2\sqrt{\alpha}$
%for $\alpha>0$, and  $0\leq\xi\ <\infty$ for $\alpha<0$.
%
From  \eq{presion} we can write the pressure as%
 \beq\fl
\lab{presion1}{p}(\xi)=\frac{C}{8\pi} \left(
 \frac{C_3\,\sin\sqrt{\alpha} (\xi+C_2)}{\sqrt{\alpha}}\right)^{2\frac{\eta+1}{\eta +7}}\,\Bigl(\cos\sqrt{\alpha}
(\xi+C_2)\Bigr)^{2\frac{\eta+1}{\eta-1}}\,=\frac{
C}{8\pi}\,g^{\xi\xi}\,. \eeq

Now, we define a new variable $u$ such that\beq u=
\frac{C_3^2\,\sin^2\sqrt{\alpha} (\xi+C_2)}{\alpha}, \eeq and,
without
losing generality for $\eta \neq -7$ , we set $C_1=C_3=\frac{\eta+7}{6}\,\kappa$, where $\kappa$ is a new arbitrary constant. %
In terms of $u$, the metric \eq{met1} becomes%
\beqa \lab{met3}  ds^2=-
u^{-\frac{2}{\eta +7}}\,\left(1-\beta
u\right)^{-\frac{2}{\eta-1}}\, dt^2 + u^{\frac{4}{\eta
 +7}}\left(dx^2+dy^2\right)\nn +\frac{9}{(\eta+7)^2\kappa^2}\,u^{-2\frac{\eta+4}{\eta
+7}}\,\left(1-\beta
u\right)^{-\frac{2 \eta}{\eta-1}}\, du^2,\eeqa%
where%
\beq\lab{beta}\beta=\frac{{9\, C\,(\eta-1)}}{4(\eta+7)\,\kappa^2}.\eeq%
The range of the coordinate $u$ is $0\leq u\leq
 1/\beta$ for $\beta>0$, and  $0\leq u <\infty$ for
$\beta<0$.  The transformation $f=u^{\frac{2}{\eta
 +7}}$ brings it to the form given by Collins in Ref. \cite{collins} \footnote{Notice that there are some typos in the expression for $g$ in page 2274 of Ref.\cite{collins}, the correct one is $g=g_0[2BM_0af^{a-3}/(1+Bbf^a)^{1-1/b}]^{-(\gamma-1)/\gamma}$.}.%

This family of solutions depending on three  parameters $C$,
$\eta$, and $\kappa$ corresponds to a \st\, full with a  static plane symmetric
perfect fluid   satisfying the equation of
state $\rho=\eta\, p$.

Notice that as $C\rightarrow0$,  i.e., the vacuum limit,
$\beta\rightarrow0$ and then \eq{met3} becomes
\beqa \lab{met4} ds^2=-  u^{-\frac{2}{\eta +7}}\,
 dt^2 + u^{\frac{4}{\eta
 +7}}\left(dx^2+dy^2\right)+\frac{9}{(\eta+7)^2 \kappa^2}\,u^{-2\frac{\eta+4}{\eta
+7}}\, du^2. \eeqa%
Setting \beq 1-\kappa\hat{z}= u^{\frac{3}{\eta+7}},\eeq%
we get%
\beqa \lab{met5} ds^2=-
\left(1-\kappa\hat{z}\right)^{-\frac{2}{3}}\,
 dt^2 + \left(1-\kappa\hat{z}\right)^{\frac{4}{3}}\left(dx^2+dy^2\right) +
 d\hat{z}^2, \eeqa%
which is Taubs's vacuum plane solution \cite{taub} in the coordinates
 used in \cite{gs}.

Since we are here interested in  ordinary matter ( i.e.,
$\rho>p>0$ and so $C>0$ and $\eta>1$), we will analyze the solution for
$\eta>1$. However,  \eq{met3} has a much wider range of validity.
In fact, it is valid for every value of $C$ and every
$\eta\neq-7$. For {\em abnormal } matter some interesting
solutions arise. However, the complete analysis turns out to be
somehow involved. So, for the sake of clarity, we leave the
complete study to a forthcoming publication \cite{gs2}, and
restrict ourselves here to the case $\eta>1$ and $C>0$. In this
case, \eq{beta} shows that $\beta>0$.
\subsection{The physical vertical distance coordinate $\hat{z}$}
The physical  distance from the ``plane" where $u=0$ to that
sitting at  $u$ is %
\beqa \int_0^{u}\sqrt{g_{\mch uu}}\, du'
=\frac{3}{(\eta+7)\kappa}\,\int_0^{u}\,u'^{\frac{3}{\eta
+7}-1}\,\left(1-\beta u'\right)^{-\frac{\eta}{\eta-1}}\, du'\nn %=
%\frac{3}{(\eta+7)\kappa}\, \beta^{-\frac{3}{\eta+7}}B_{\beta
%u}\left(\frac{3}{\eta+7},-\frac{1}{\eta-1}\right) \nn
=\frac{1}{\kappa}\,u^{\frac{3}{\eta+7}}\,_2F_1\!\left(\frac{3}{\eta+7},\frac{\eta}{\eta-1};1+\frac{3}{\eta+7};\beta
u\right)\nn \fl
=\frac{1}{\kappa}\,u^{\frac{3}{\eta+7}}\,\left(1-\beta\,
u\right)^{-\frac{1}{\eta-1}}\,\,_2F_1\!\left(1,1+\frac{3}{\eta+7}-\frac{\eta}{\eta-1};1+\frac{3}{\eta+7};\beta u\right), \eeqa %
where $_2F_1(a,b;c;z)$ is the Gauss hypergeometric function (see for
example \cite{tablarusa,abra}). The hypergeometric function in the
last line is finite for $0\leq u\leq 1/\beta$, since $c-a-b=
\frac{1}{\eta-1}>0$; furthermore \cite{abra}
\beqa \fl
_2F_1\!\left(1,1+\frac{3}{\eta+7}-\frac{\eta}{\eta-1};1+\frac{3}{\eta+7};1\right)=
\frac{\Gamma(1+\frac{3}{\eta+7})\,\Gamma(\frac{1}{\eta-1})}{\Gamma(\frac{3}{\eta+7})\,\Gamma(\frac{\eta}{\eta-1})}=\frac{3(\eta-1)}{\eta+7}. \eeqa %
Hence, we see that as $u$ goes from $0$ to $1/\beta$, the physical
distance goes from $0$ to $\infty$.

Thus, we can write %
\beq\lab{hatz}1-\kappa\hat{z}=
u^{\frac{3}{\eta+7}}\,_2F_1\!\left(\frac{3}{\eta+7},\frac{\eta}{\eta-1};1+\frac{3}{\eta+7};\beta u\right), \eeq%
and, in terms of the coordinate $\hat{z}$, \eq{met3} becomes%
\beqa \lab{met6}\fl ds^2=-  u^{-\frac{2}{\eta +7}}\,\left(1-\beta
u\right)^{-\frac{2}{\eta-1}}\,
 dt^2 + u^{\frac{4}{\eta
 +7}}\left(dx^2+dy^2\right) + d\hat{z}^2,\nn\fl
-\infty<t<\infty,\quad\,-\infty<x<\infty,\quad\,-\infty<y<\infty,\quad\,-\infty<\hat{z}<1/\kappa\,.\eeqa%
 where  $u$  ($0\leq u\leq1/\beta$) is implicitly
given in terms
 of $\hat{z}$ through \eq{hatz}.

For the sake of clarity, we present here an explicit example where the above expressions can be written in terms of algebraic
functions.
If $\eta= 5 $, we can write \eq{hatz} as%
\beqa 1-\kappa \hz=u^{\frac{1}{4}}\,\left(1-\beta\,
u\right)^{-\frac{1}{4}}\,\,_2F_1\!\left(1,0;\frac{5}{4};\beta
u\right)= {\left(\frac{u}{1-\beta
u}\right)}^{\frac{1}{4}}\,, \eeqa %
and so \beq u=\frac{(1-\kappa \hz)^{{4}}}{1+\beta(1-\kappa
\hz)^4}\,. \eeq
Therefore, for $\eta= 5 $, \eq{met6} becomes%
\beqa\fl \lab{met7} ds^2=- \frac{\left(1+\beta(1-\kappa
\hz)^4\right)^{\frac{2}{3}}}{(1-\kappa \hz)^{\frac{2}{3}}} \,
 dt^2 + \frac{(1-\kappa \hz)^{\frac{4}{3}}}{\left(1+\beta(1-\kappa
\hz)^4\right)^{\frac{1}{3}}}\left(dx^2+dy^2\right) + d\hat{z}^2,\nn\fl
-\infty<t<\infty,\quad\,-\infty<x<\infty,\quad\,-\infty<y<\infty,\quad\,-\infty<\hat{z}<1/\kappa\,.\eeqa%

\subsection{The singularity}

We see that the metric has a \st\, curvature singularity at
$\hz=1/\kappa$ ($u=0$), as it occurs for   Taub's empty plane
symmetric \st\,. This singularity, as it happens with the
Schwarzschild's one at $r=0$, is a {\em true} one in the sense
that \st\ curvature diverges. For straightforward computation of
the scalar quadratic in the
Riemann tensor yields%

  \beqa\lab{RR}\fl R_{abcd}R^{abcd}=\frac{64}{27}\, {\kappa}^4\,u^
     {-\frac{12}{\eta + 7}}\,(1-\beta u)^
     {2\frac{ \eta+1  }{\eta -1 }}\, \Bigl( 1+
      \frac{2(\eta + 1 )
(\eta + 3)}{3(\eta -1 )}\,\,\beta u\Bigr.\nn  +\Bigl.
    \frac{ 159 + 172\,\eta  + 142\,{\eta }^2 + 36\,{\eta }^3 +
  3\,{\eta }^4 }{12( \eta -1)^2}\,\beta^2 u^2\Bigr)\ , \eeqa
so $ R_{abcd}R^{abcd}\rightarrow\infty$ as $u\rightarrow0$  ($\hz
\rightarrow1/\kappa$).

 On the other hand, $ R_{abcd}R^{abcd}\rightarrow0$ as $u\rightarrow1/\beta$ ($\hz
\rightarrow-\infty$ ) since $\frac{2\,\left( \eta+1 \right) }{\eta
-1 }>0$, suggesting it is asymptotically flat at spatial infinite
in the $-\hz$ direction.

 Note that each slice of \st \  $t=t_0$ and $u=u_0$ is a Euclidean
plane with metrics \beq  d\ell^2=u^{\frac{4}{\eta
+7}}\left(dx^2+dy^2\right) ,\eeq so its ``size''  contracts when
$\hz$ increases  and becomes a point at the singularity
$\hz=1/\kappa$ ($u=0$).

Notice that, beyond the singularity (i.e., $\hz
\rightarrow1/\kappa$), a mirror
copy of the  \st\  emerges, as it occurs at the vertex of a cone.
But, as we shall show below, no geodesic can  go from
one to the other.

\subsection{The pressure}

From \eq{presion1}, the pressure can be written as \beq
\lab{presion2}{p}(u)=\frac{ C}{8\pi}\, u^{\frac{\eta+1}{\eta
+7}}\,\left(1-\beta u\right)^{\frac{\eta+1}{\eta-1}}, \eeq so
$p(0)=p(1/\beta)=0$. It follows readily from \eq{hatz}  that it
vanishes like $(1-\kappa \hz)^{\frac{\eta+1}{3}}$ as
$\hz\rightarrow1/\kappa$ ($u=0$), and like $(1-\kappa
\hz)^{-(\eta+1)}$ as $\hz\rightarrow-\infty$
($u\rightarrow1/\beta$). On the other hand, it has a maximum at
\beq
u=\frac{(\eta-1)}{2\,\beta\,(\eta+3)}\,=\frac{2\,\kappa^2\,(\eta+7)}{9\,C\,(\eta+3)}\,.\eeq
In Fig. 1, we show a plot of $p(\hz )$.

\begin{figure}[t]\label{figura}\begin{center}
\includegraphics[height=5.5cm]{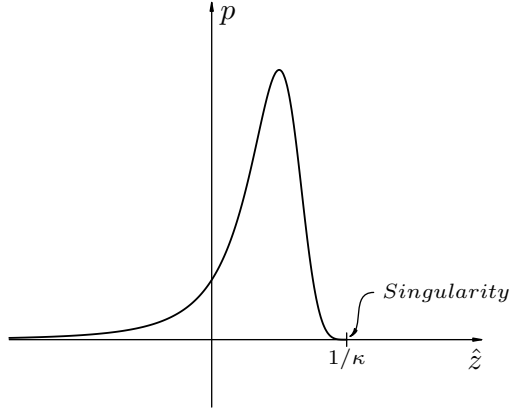}
\caption{The pressure. }\end{center}\end{figure}

Since $p=0$ (and $\rho=0$) at the singularity, so $T_{ab}=0$ there,
we see that it is an {\em empty}  (without matter) singularity.
%Notice that, the vanishing of $T_{ab}=0$  implies the vanishing of $G_{ab}$, $R$ and $R_{ab}$ at %the singularity, and so it is the Weyl tensor which diverges.
Thus,  the attraction
of the infinite amount of  ordinary matter curves \st\, in such a way that a
singularity arises in a  place free of matter.

\section{The geodesics }

We want to study the geodesics in this \st . Since the metric is
independent of $t$, $x$ and $y$, the momentum covector components
$p_t$, $p_x$ and $p_y$ are constant along the geodesics. For
timelike geodesics, we choose $\tau$ to be the proper time; and
for null ones, we choose $\tau$ to be an affine parameter. So, we
can write \beqa \fl
\left(\frac{du}{d\tau}\right)^2=\frac{(\eta+7)^2\kappa^2}{9}\,u^{-\frac{6}{\eta
+7}+2} \left(1-\beta u\right)^{\frac{2\eta}{\eta-1}}\,
\nn \times\left[u^{\frac{2}{\eta +7}}\,\left(1-\beta
u\right)^{\frac{2}{\eta-1}}\,{\tilde{E}}^2 -\mu-
u^{-\frac{4}{\eta +7}}\, {(\tilde{p_x}^2+\tilde{p_y}^2)}\right],\label{rebote}\\
\frac{dt}{d\tau}=u^{\frac{2}{\eta +7}}\,\left(1-\beta
u\right)^{\frac{2}{\eta-1}}\,
{\tilde{E}},\lab{dt}\\\frac{dx}{d\tau}=u^{-\frac{4}{\eta
+7}}\,{\tilde{p}_x},\\\frac{dy}{d\tau}=u^{-\frac{4}{\eta +7}}\,
 {\tilde{p}_y}, \eeqa  where $\mu=1$, ${\tilde{E}}=-p_t/m$,
$\tilde{p_x}=p_x/m$ and $\tilde{p_y}=p_y/m$ for timelike
geodesics; and $\mu=0$, ${\tilde{E}}=-p_t$, $\tilde{p_x}=p_x$ and
$\tilde{p_y}=p_y$ for null  ones.

The right hand side of (\ref{rebote}) cannot be negative.  Thus,%
\beq \lab{rebote2} u^{\frac{6}{\eta +7}}\,\left(1-\beta
u\right)^{\frac{2}{\eta-1}}\,{\tilde{E}}^2 -u^{\frac{4}{\eta
+7}}\,\mu- \, {(\tilde{p_x}^2+\tilde{p_y}^2)} >0\,. \eeq %
Therefore, only vertical null geodesics  touch   the singularity at $u=0$ ($\hz
\rightarrow1/\kappa$) and
bounce following its travel to $u=1/\beta$ ($\hz=
-\infty$). Whereas, as it is shown below, non-vertical null ones as well as massive particles bounce
before getting to it and bounce again before reaching $\hz=-\infty$.

 For the former
case, the geodesic equation can be
  integrated in  closed form. Indeed, when $\mu=0$ and
$\tilde{p_x}^2+\tilde{p_y}^2=0$,  we get from \eq{rebote} and
 \eq{dt}
\beqa \frac{dt}{du}&=\pm
&\frac{3}{(\eta+7)\kappa}\,u^{\frac{4}{\eta +7}-1}\,\left(1-\beta
u\right)^{-1}\,.  \eeqa So \beqa |t-t_0|=
\frac{3}{4\kappa}\,u^{\frac{4}{\eta +7}}\,_2F_1\!\left(\frac{4}{\eta
+7},1;1+\frac{4}{\eta +7};\beta u\right). \eeqa
   Now, taking into account that when $z\rightarrow1$,
\beq\lab{2F1}
_2F_1\!\left(a,b;a+b;z\right)\rightarrow-\frac{\Gamma(a+b)}
{\Gamma(a)\Gamma(b)}\ln\left({1-z}\right)\,, \eeq%
we see that
 \beqa
|t-t_0|\sim \cases{\frac{3}{4\kappa}\,u^{\frac{4}{\eta
+7}}\rightarrow0&\text{as}\,\,\, $u\rightarrow0$\\
-\frac{3\ \beta^{-\frac{4}{\eta
+7}}}{(\eta+7)\kappa}\, \,\ln(1-\beta u
)\rightarrow\infty&\text{as}\,\,\, $u\rightarrow1/\beta$\,.
} \eeqa

 Therefore, we can fill the whole \st\ with
never-stopping future-oriented vertical null geodesics, all of them
starting  and finishing at $\hz=-\infty$.

It follows from (\ref{rebote2}) that the movement of non-vertical photons or massive particles is constrained to the region where ${\tilde{E}}^2>\mathcal{V}(u)$, being
\beqa \mathcal{V}(u)= \frac{  u^{\frac{4}{\eta+7}}\,\mu\,+ \, (\tilde{p_x}^2+\tilde{p_y}^2)}    {{u^{ \frac{6}{\eta+7 }} \left( 1 - \beta u \right) }^{\frac{2}{ \eta-1 }}}\ . \eeqa
Clearly $\mathcal{V}(u)$ is a positive continuous function of $u$ for $0<u<1/\beta$ and, since  $\eta>1$, $V(u)\rightarrow+\infty$ when $u\rightarrow0$ or  $u\rightarrow1/\beta$. Moreover we can write

\beqa \mathcal{V}''(u)= 2\ \frac{  \left(u^{\frac{4}{\eta+7}}\,\mu\, P_1(\beta u )+ \, (\tilde{p_x}^2+\tilde{p_y}^2)\, P_2(\beta u ) \right)}    {{u^{ \frac{6}{\eta+7 }+2} ( 1 - \beta u ) }^{\frac{2}{ \eta-1 }+2}\,{\left(  \eta  -1 \right) }^2\,
    {\left(  \eta +7 \right) }^2}\ , \eeqa
where $P_1(z )$ and $P_2(z )$ are the  second degree polynomials in $z$:
\beqa \fl P_1(z )= 2\,z \Bigl( 5 + \eta \,(\eta+10 )\Bigr)\Bigl(z (\eta +3)-( \eta -1)\Bigr) +\, ( \eta -1)^2\,(\eta+9)\ , \\  \fl P_2(z )= 2\,z \Bigl( 1 + \eta \,( \eta+14 )\Bigr)\Bigl(2 z (\eta +1)-3(\eta -1)\Bigr) +\, 3\ ( \eta -1)^2\,(\eta+13)\ . \eeqa
We can readily see that both polynomials   are positive definite  in $0<z<1$  for  $\eta>1$. Then $\mathcal{V}''(u)>0$, and so $\mathcal{V}(u)$ has one and only one minimum $\mathcal{V}_0$. Therefore for any value of ${\tilde{E}}^2$ greater than $\mathcal{V}_0$ we find two (and only two) turning points where ${\tilde{E}}^2=\mathcal{V}(u)$. Thus timelike geodesics as well as non-vertical null ones oscillate between two planes determined by  initial conditions.

As no geodesic can cross the singularity we can think of it by removing the point
$\hz=1/\kappa$ as the boundary of the \st.

Hence, the attraction of the {\em infinite} amount of matter filling the spacetime with plane symmetry shrinks the
\st\ in such a way that it finishes  at the empty singular
boundary.

\section{Concluding remarks }

We have done a detailed study of the Collins  exact solution of Einstein's equations
corresponding to a static and plane symmetric  distribution of
 ordinary matter with density proportional to the pressure.

This  simple spacetime turns out to present some  somehow astonishing properties. Namely, the attraction of the {\em infinite} amount of matter filling the spacetime with plane symmetry shrinks the \st\ in such a way that it finishes  at an empty singular boundary.

Only vertical null geodesics  touch   the boundary and
bounce following  its travel to $-\infty$, whereas non-vertical null ones as well as massive particles bounce
before getting to it and oscillate between two planes.

Since the density vanishes at the singularity this example shows
that,  as  suggested in \cite{gs}, repelling singularities can
arise in a place free of matter.

For  {\em abnormal } matter ({\em i.e.} $\eta<1$ or $C<0$) some
interesting solutions arise, depending on the combination of the
parameters $C$ and $\eta$. For instance, for  $C<0$ and $\eta>1$
the \st\ is bounded between two singularities of this kind
separated by a finite distance \cite{gs2}.

\section*{References}
%%%%%%%%%%%%%

\end{document}